\documentclass[preprint,showpacs,preprintnumbers,amsmath,amssymb]{revtex4}

\usepackage{graphicx}

\begin{document}

\title{
Gravitational waveforms for 2- and 3-body gravitating systems 
}
\author{Yuji Torigoe}
\author{Keisuke Hattori}
\author{Hideki Asada} 
\affiliation{
Faculty of Science and Technology, Hirosaki University,
Hirosaki 036-8561, Japan} 

\date{\today}

\begin{abstract}
Different numbers of self-gravitating particles  
(in different types of periodic motion) are most likely 
to generate very different shapes of gravitational waves, 
some of which, however, can be accidentally almost the same. 
One such example is a binary and a three-body system for Lagrange's solution. 
To track the evolution of these similar waveforms, 
we define a chirp mass to the triple system. 
Thereby, we show that the quadrupole waveforms cannot 
distinguish the sources. 
It is suggested that waveforms with higher $\ell$-th multipoles 
will be important for classification of them 
(with a conjecture of $\ell \leq N$ for $N$ particles). 
\end{abstract}


\maketitle

\noindent \emph{Introduction.--- } 
Can one see an apple fall at dark night? 
This is an inverse problem in gravitational waves astronomy. 
It can be specifically stated as  
``how can we know the source information 
such as the number of particles, their geometrical shape and 
motion from observations of gravitational waves?'' 
This problem is analogous to the well-known one for the sound, 
which was raised by Kac in his celebrated paper \cite{Kac} entitled 
``{\it Can one hear the shape of a drum?}''  
Seeking an answer is beyond the scope of this paper. 
As a specific issue which is related with the inverse problem, 
we shall examine gravitational radiation by a certain class  
of orbital motion of self-gravitating objects. 

In the near future, direct detections of gravitational ripples 
(and consequently gravitational waves astronomy) 
will come true owing to a lot of efforts by 
the on-going or designed detectors 
\cite{Centrella,Tinto,Mioa,Miob,LIGO-GRB,TAMA}. 
One of the most promising astrophysical sources is 
inspiraling and finally merging binary compact stars. 
Numerical relativity has succeeded in 
simulating merging neutron stars and black holes 
\cite{Shibata,Pretorius,BBH06a,BBH06b,BBH06c}. 
Analytic methods also have nicely 
prepared accurate waveform templates for 
inspiraling compact binaries, 
notably by the post-Newtonian approach 
(See \cite{Blanchet,FI} for reviews) 
and also by the black hole perturbations 
especially at the linear order in mass ratio  
(See also \cite{ST} for reviews). 
Bridges between the inspiraling stage and the final merging phase 
are currently under construction (e.g., \cite{DN08a,DN08b}). 

\begin{figure}[t]
\includegraphics[width=8cm]{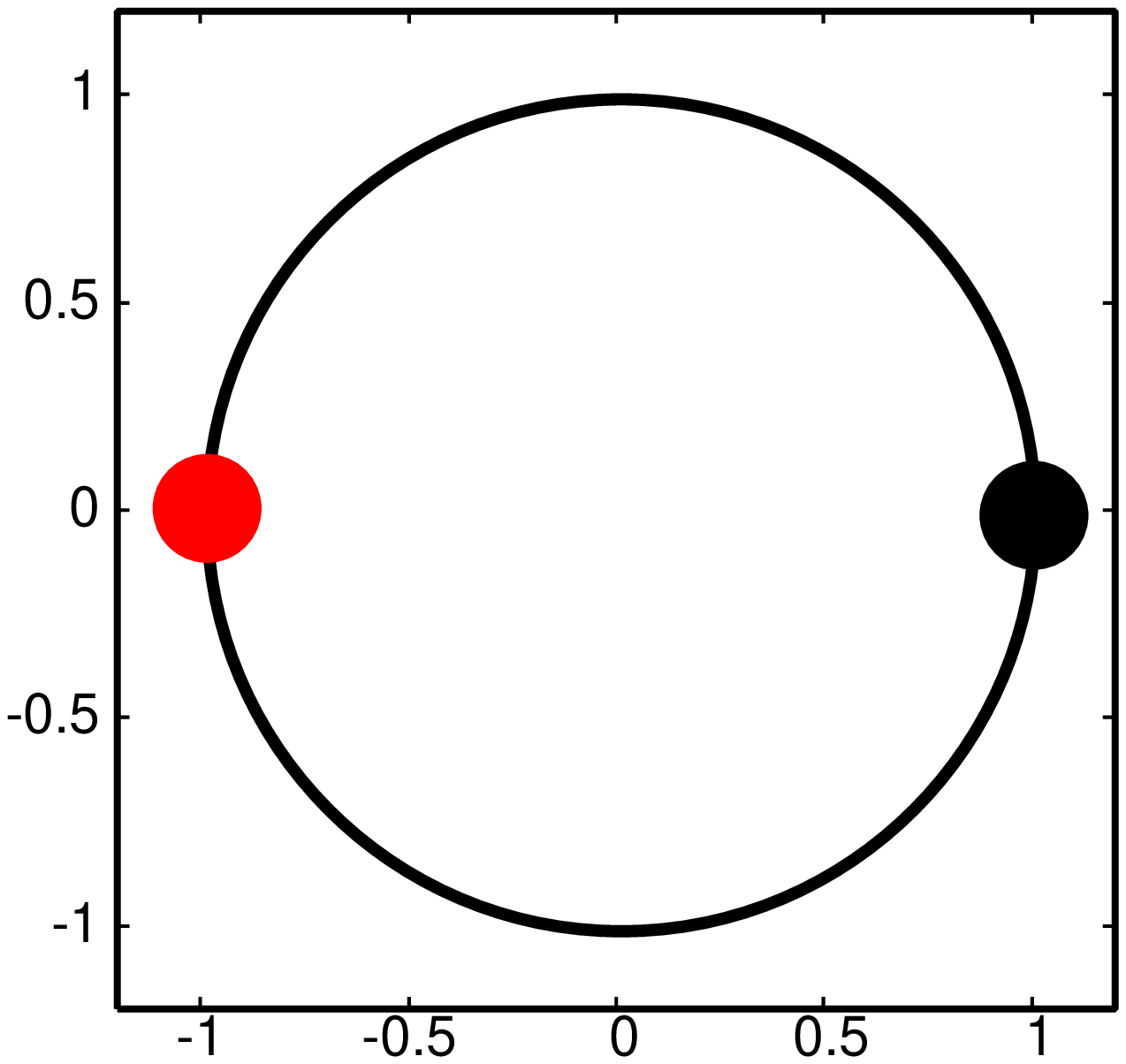}
\includegraphics[width=8cm]{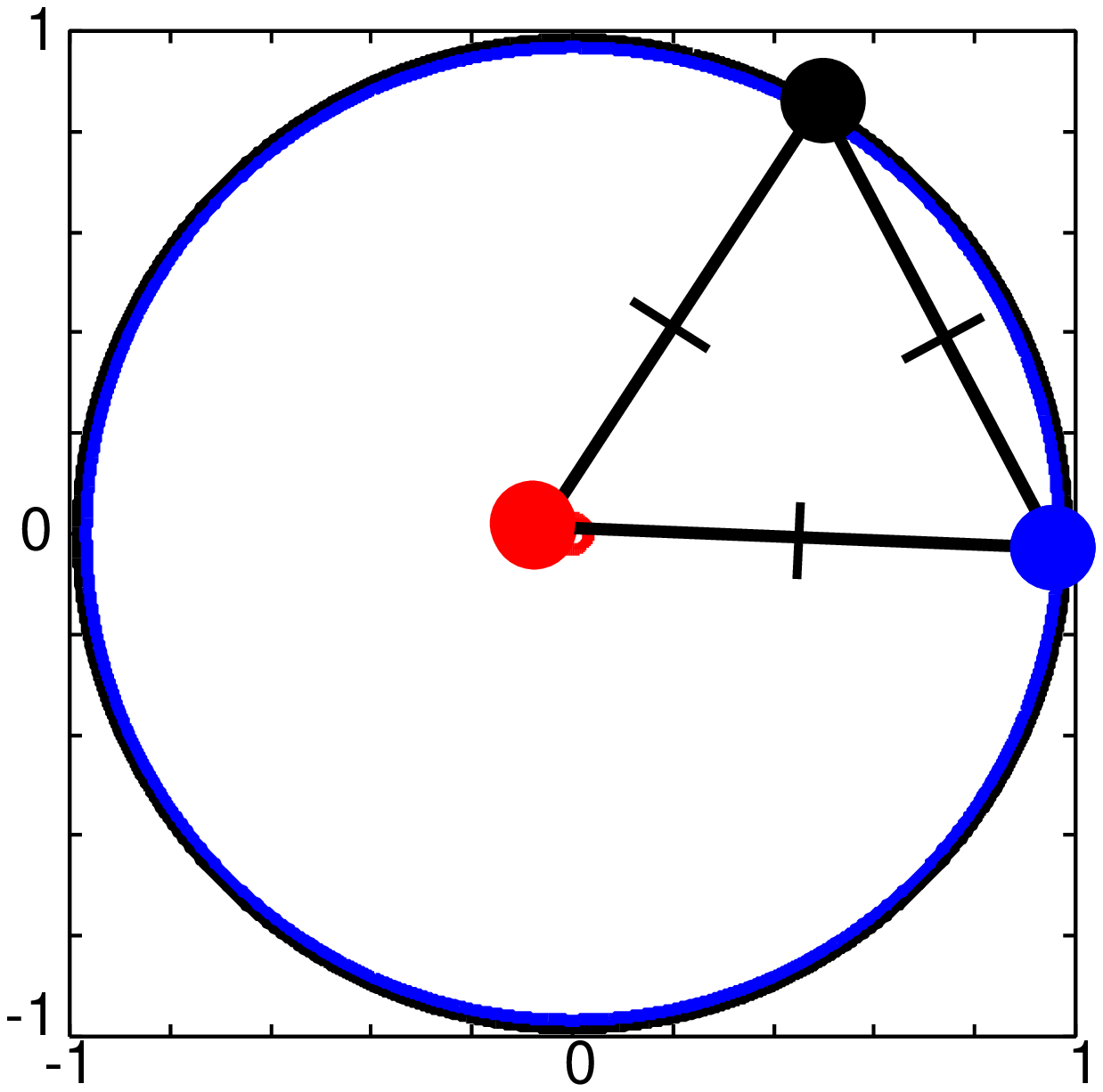}
\\
\includegraphics[width=8cm]{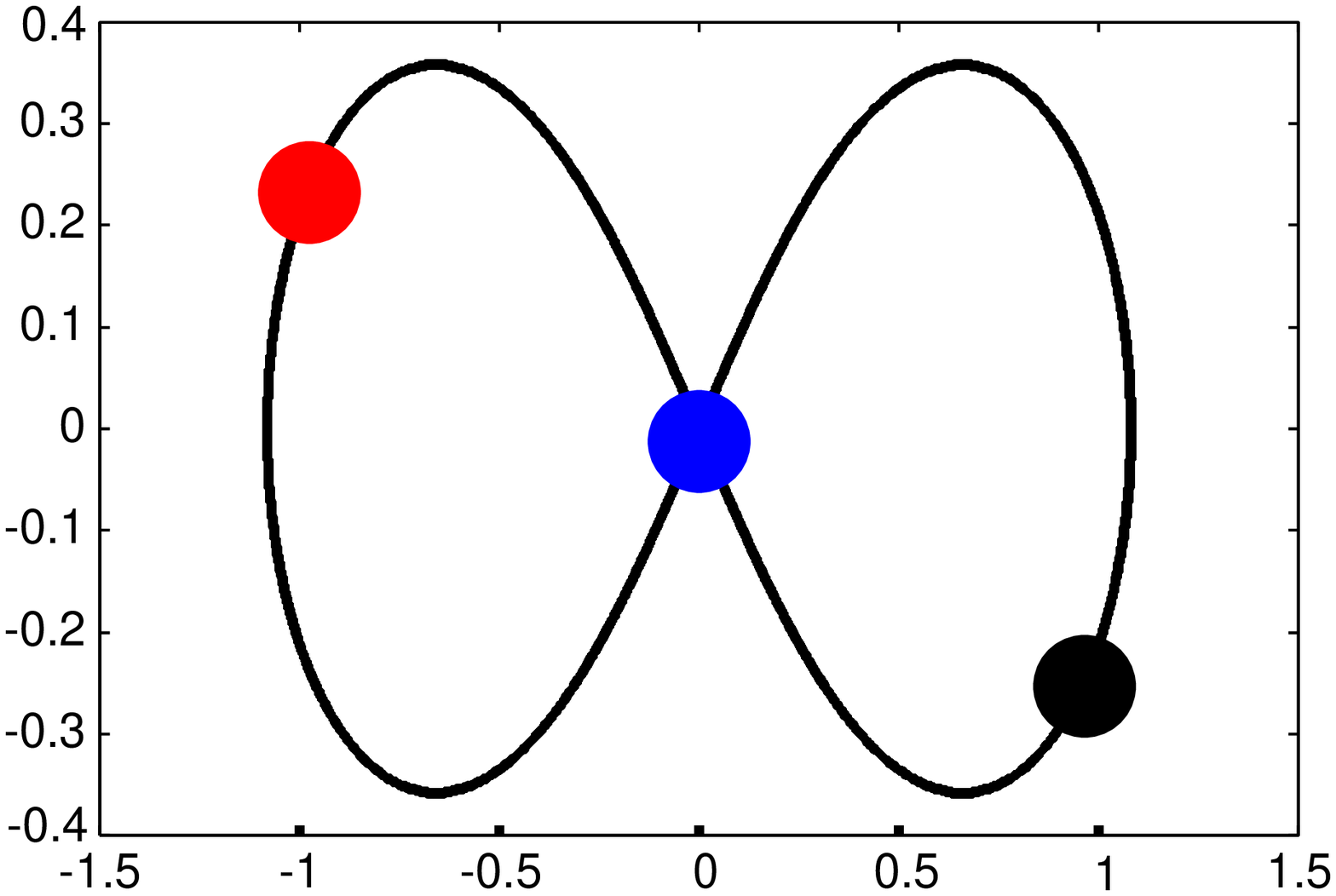}
\includegraphics[width=8cm]{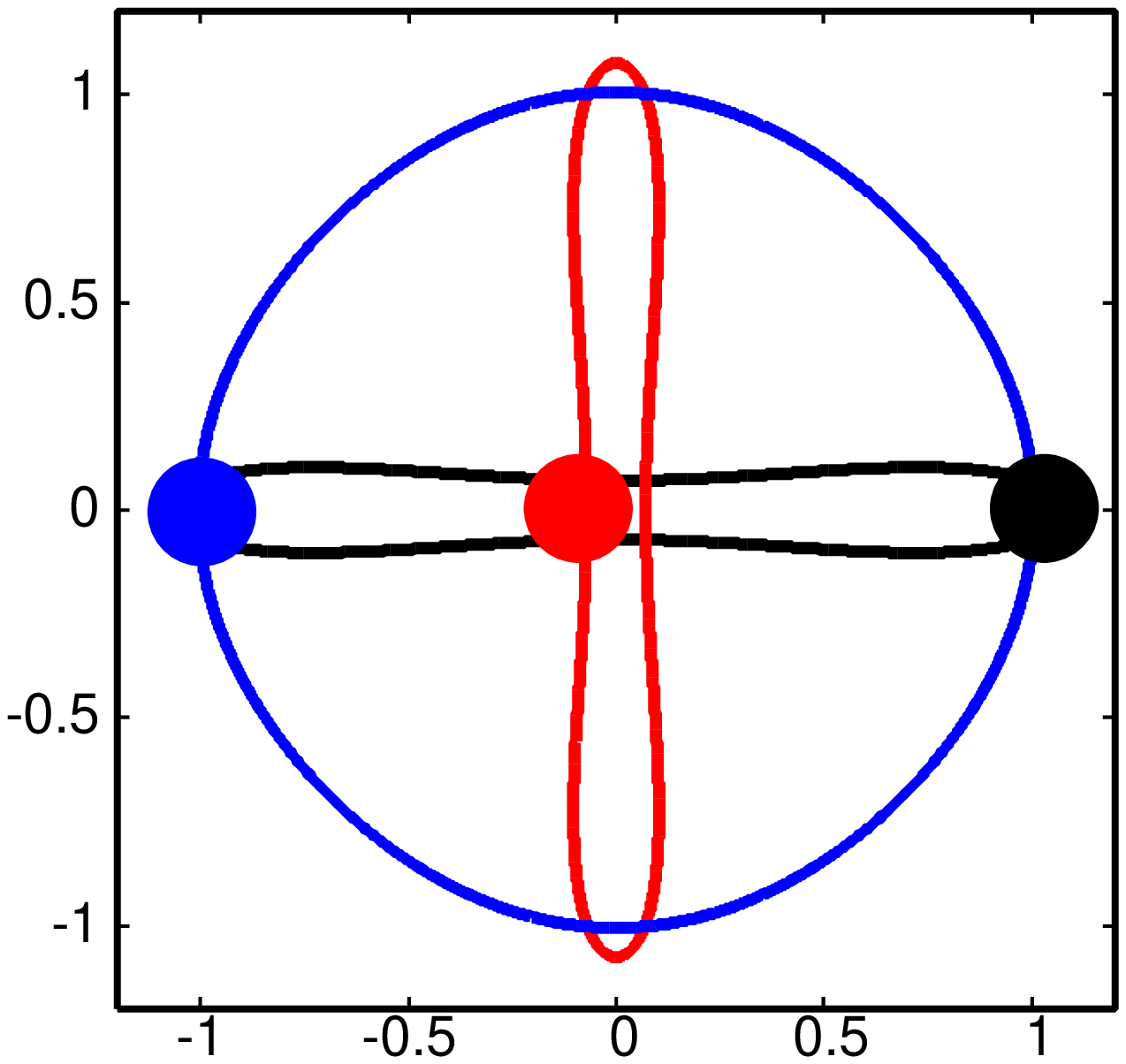}
\caption{
Orbital shapes. 
(a) Top left: Circular orbit for two-body system 
as a reference. 
(b) Top right: Triangle solution by Lagrange.
(c) Bottom right: Criss-cross orbit by Henon.  
(d) Bottom left: Figure-eight trajectory by Moore.  
}
\label{f1}
\end{figure}

There is a growing interest in potential astrophysical
sources of gravitational waves involving 3-body interactions 
(e.g., \cite{CIA,LN} and references therein). 
It is well-known that even the classical three-body (or N-body) 
problem in Newtonian gravity admits an increasing number of solutions 
\cite{Danby,Marchal}. 
Some of the orbits are regular, while the others are chaotic.
For simplicity, we focus on several periodic orbits 
of three body system; Lagrange's triangle, Henon's criss-cross 
and Moore's figure-eight, which are explained later 
(See also Fig. $\ref{f1}$).  
Here, it should be noted that 
Nakamura and Oohara \cite{NO} studied numerically the luminosity of 
gravitational radiation by N test particles 
orbiting around a Schwarzshild black hole, 
as an extension of Detweiler's analysis of the $N=1$ case 
\cite{Detweiler} 
by using Teukolsky equation \cite{Teukolsky}, 
in order to show the phase cancellation effect, 
which had been pointed out by Nakamura and Sasaki \cite{NS}. 
Their N particles are {\it test} masses but not self-gravitating. 
Another inverse problem of reconstructing the gravitational wave 
signal from the noisy data acquired by a network of detectors has been 
discussed (e.g., \cite{GT,Nicholson}). 
Our aim and setting are completely different from those 
of the existing works. 

The purpose of this paper is (1) to point out a case where 
very similar shapes of waves are generated accidentally 
by different numbers of particles 
and 
(2) to show that the usage of higher multipole contributions 
will be necessary for distinguishing such sources. 
In order to track the evolution of the waveforms, 
we shall define the chirp mass 
so as to extend to a three-body system. 
Thereby, we shall show that the octupole order 
is required to disentangle such very similar waveforms 
that coincide with each other 
at the quadrupole level. 
This will suggest that theoretical waveforms 
including sufficiently higher $\ell$-th order multipole  
will be important for classification of sources generating 
such similar waveforms (with a conjecture about $\ell$ and $N$). 

Throughout this paper, we take the units of $G=c=1$. 

\noindent \emph{Some periodic orbits for three-body systems.---} 
For simplicity, we assume that the motion of massive bodies 
follows the Newtonian equation of motion. 
It is impossible to describe all the solutions 
to the three-body problem even for the $1/r$ potential, 
as mentioned above. 
The simplest periodic solutions for this problem 
were discovered by Euler (1765) and by Lagrange (1772). 
The Euler's solution is a collinear solution, 
in which the masses are collinear at every instant 
with the same ratios of their distances. 
The Lagrange's one is an equilateral triangle solution 
in which each mass moves in an ellipse in such a way 
that the triangle formed by the three bodies revolves.  
Let us take as another interesting solution 
the so-called criss-cross orbit 
found by Henon in 1976 \cite{Henon}  
(See also \cite{MN} for the initial condition for each mass 
and recent extensions of the solution). 

Since the figure-eight solution was found first by Moore 
by topological classification \cite{Moore}, 
choreographic solutions have recently attracted 
increasing interests in astronomy, mathematics and physics,  
where a solution is called choreographic 
if every massive particles move periodically 
in a single closed orbit. 
The figure-eight solution is that three bodies 
move periodically in a single figure-eight \cite{Moore}. 
The existence of such a figure-eight orbit was proven 
by Chenciner and Montgomery \cite{CM}, 
where the numerical initial condition for each mass is also given. 
This odd solution is remarkably stable in Newtonian gravity 
\cite{Simo,GMF}. 
Heggie discussed a formation mechanism as 
an outcome from scattering of two binaries \cite{Heggie}. 
Eventually its unicity up to scaling and rotation has been 
recently proven 
\cite{Montgomery05}. 
The trick figure eight remains true even if we consider 
the general relativistic effects at the post-Newtonian order \cite{ICA} 
and also at the second post-Newtonian one \cite{LN}. 
This is a marked contrast to a binary case, which produces 
a complicated flowerlike pattern by the periastron advance 
in Einstein gravity. 
It is interesting to investigate relativistic effects 
on various kinds of orbital motions,  
which are discussed mostly in Newtonian gravity. 
It is a topic of future study. 
The radiation by the figure eight has been also investigated \cite{CIA}

\begin{figure}[t]
\includegraphics[width=8cm]{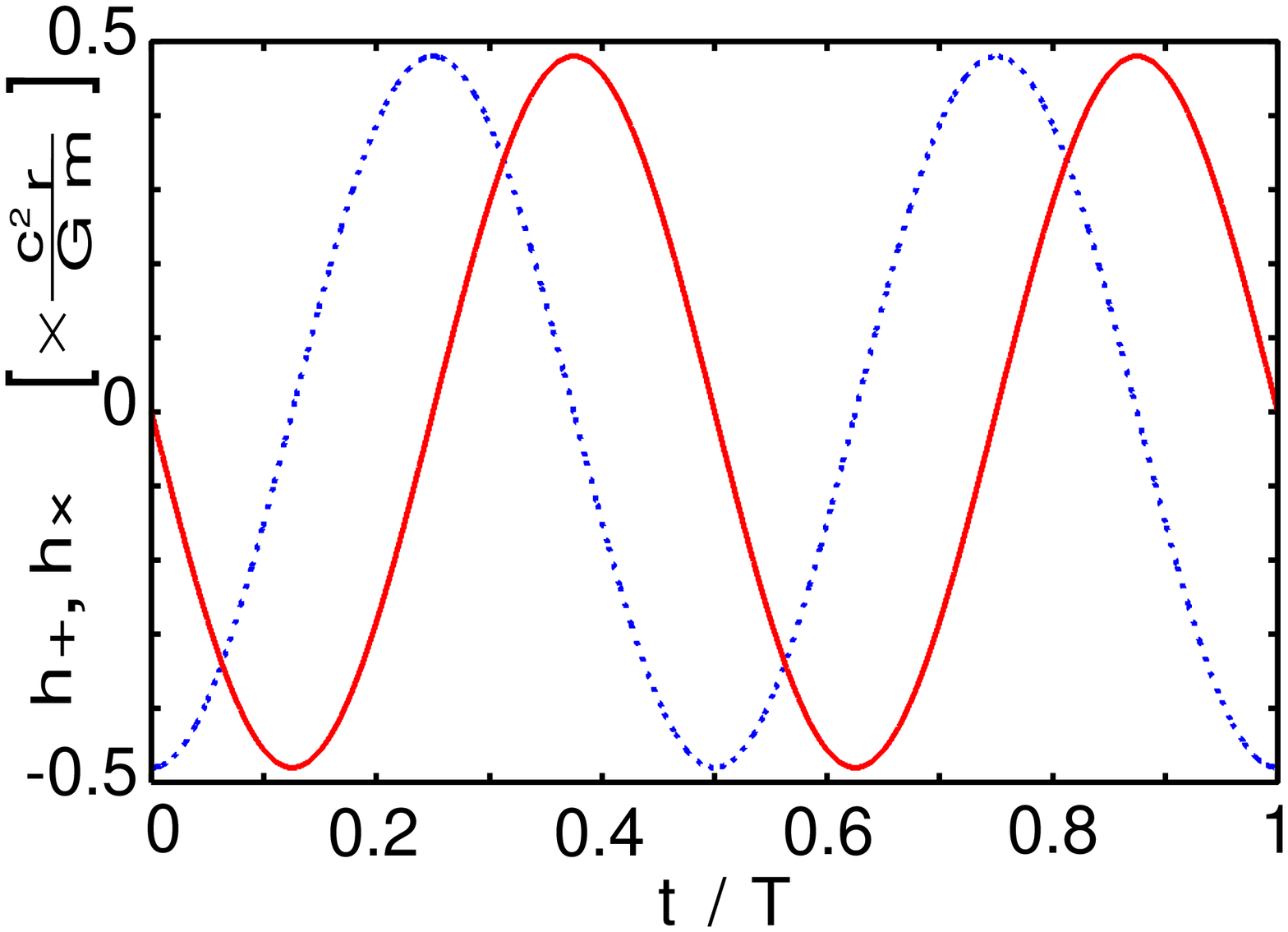}
\includegraphics[width=8cm]{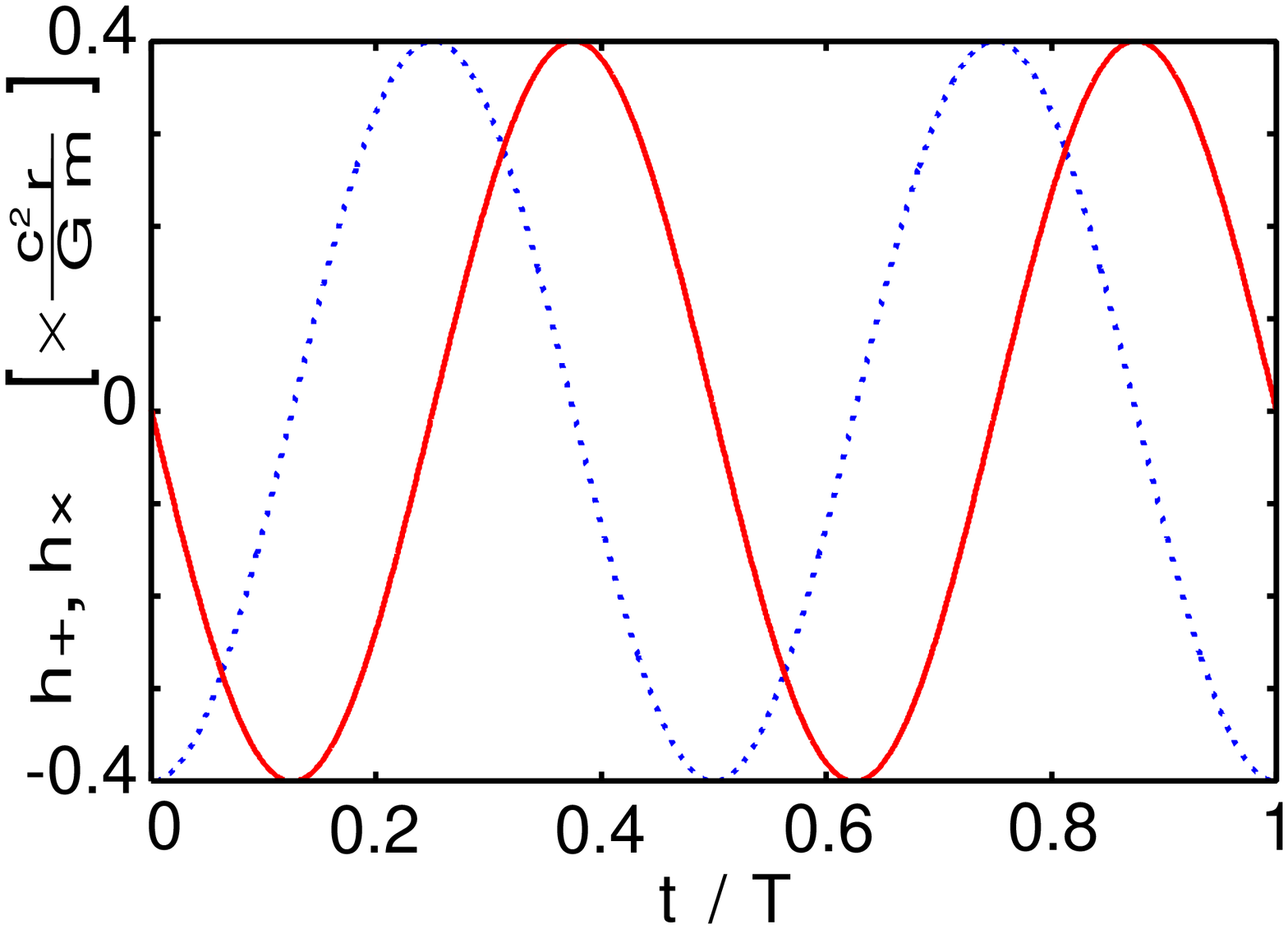}
\\
\includegraphics[width=8cm]{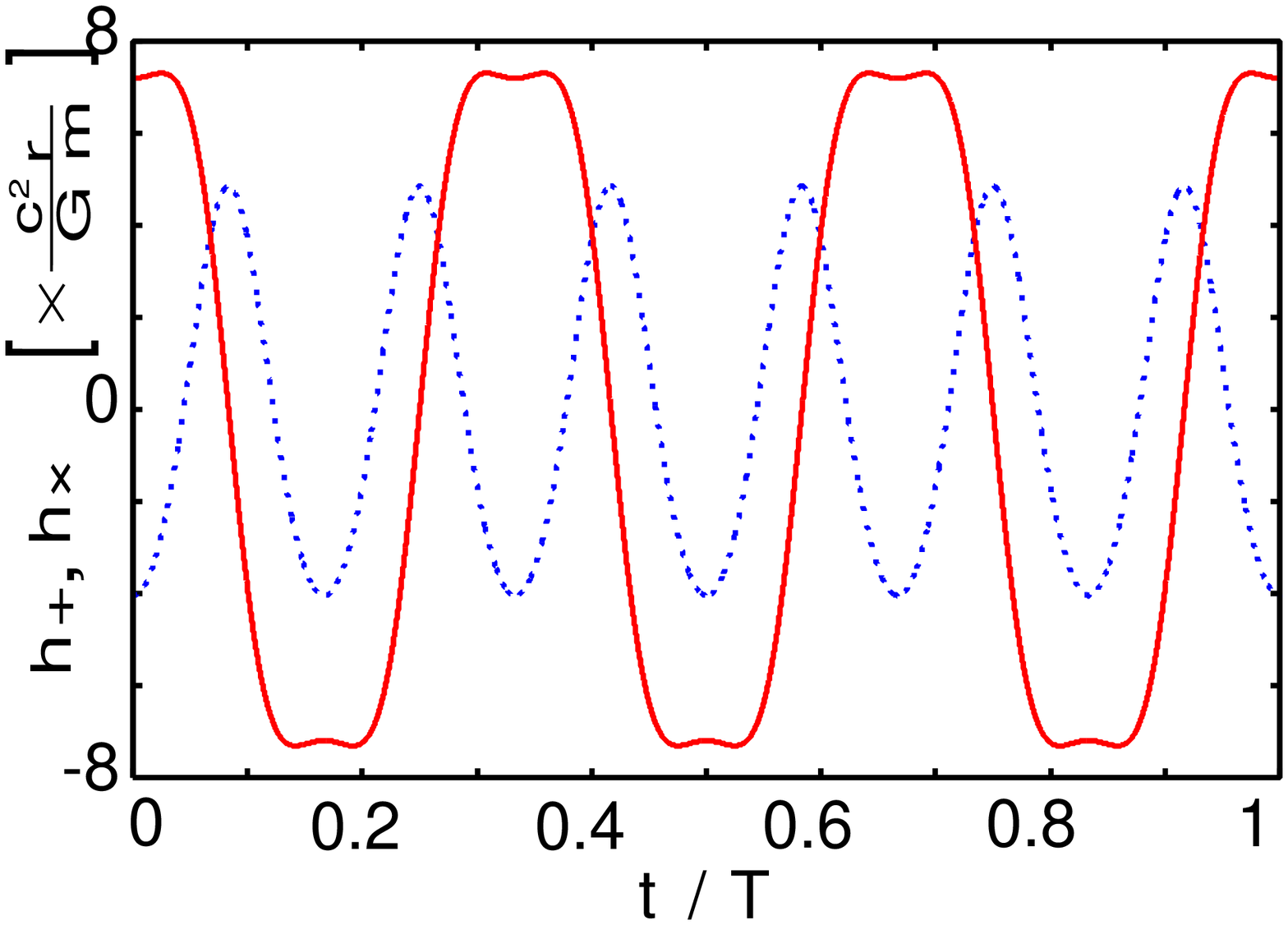}
\includegraphics[width=8cm]{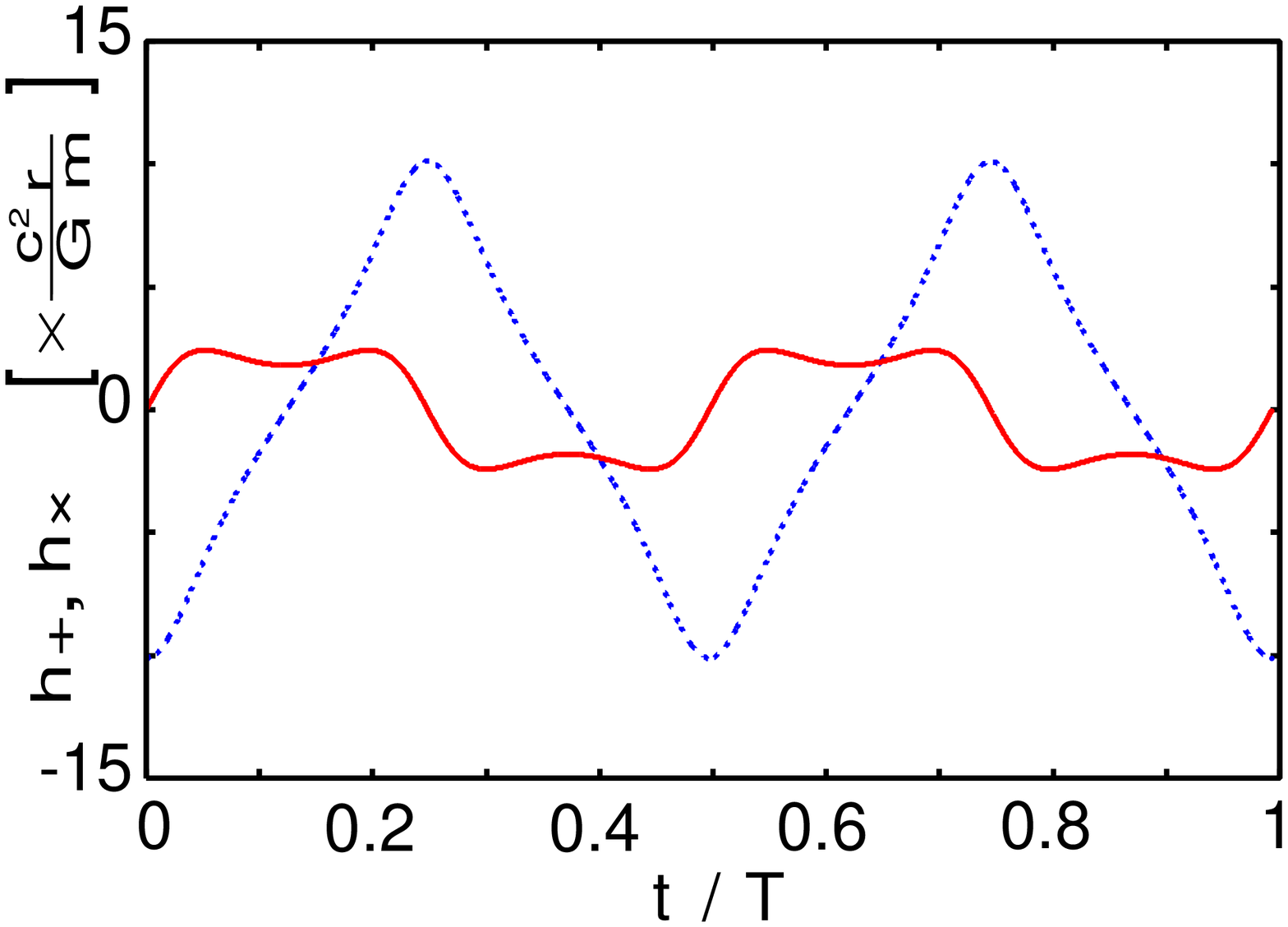}
\caption{
Gravitational waveforms in arbitrary units ($T=$ orbital period). 
Dotted blue and solid red curves denote 
$+$ and $\times$ modes, respectively. 
(a) Top left: Gravitational waveforms 
by binary system with a mass ratio of 2:3 
in circular motion. 
(b) Top right: Lagrange's triangle solution 
for a mass ratio of 1:2:3.
(c) Bottom right: Henon's criss-cross.   
(d) Bottom left:  Moore's figure-eight.  
Criss cross and figure eight have larger 
curvatures in the orbital shapes than 
Keplerian and Lagrangian orbits,  
which lead to larger acceleration of the particles 
and thus relatively stronger radiation. 
}
\label{f2}
\end{figure}

\noindent \emph{Gravitational waves.---}
In the previous 
part, 
we have mentioned several periodic solutions. 
Figure $\ref{f2}$ shows the gravitational radiation 
by massive particles in these periodic motions, 
where the quadrupole formula is used. 

Interestingly, the waveforms from a binary 
in circular motion and a three-body system constituting 
the Lagrange solution are the same in shape. 
It is worthwhile to mention that, if the third mass 
is extremely small, its contribution to the quadrupole waves 
becomes linear but not cubic in mass because its orbital radius 
is of the order of a triangle's side length, 
namely bounded from above. 
If one adjust properly distance $r$ from an observer to the source 
with the same orbital period, 
the waveforms (including the amplitudes) 
could perfectly agree with each other.  
  
\noindent \emph{Chirp mass for three-body systems.---}
The waveforms shown above are valid only in short term.  
The gravitational waves will gradually carry away 
the system's energy and angular momentum, 
and will eventually shrink the orbital size. 
Consequently, the amplitude and frequency of the waves 
will become larger and higher, respectively, with time. 
For a binary case, the frequency sweep is characterized 
by its chirp mass. 

Here, we investigate the evolution of the waveforms 
for a three-body system for the Lagrange's solution 
(on $x$-$y$ plane). 
The initial positions of each mass denoted by $m_p$ $(p=1, 2, 3)$ 
are expressed as 
$\mbox{\boldmath $x$}_1=(0, 0)$, 
$\mbox{\boldmath $x$}_2=a (\sqrt{3}/2, 1/2)$, 
and 
$\mbox{\boldmath $x$}_3=a (0, 1)$, 
where the side of a regular triangle is denoted as $a$. 
We take the coordinates such that 
the center of mass (COM) is at rest as 
$(x_{COM}, y_{COM})
= a (\sqrt{3} {\nu}_2/2, ({\nu}_2+{\nu}_3)/2)$), 
where the total mass and mass ratio are denoted 
as $m_{tot} \equiv \sum_p m_p$ and 
${\nu}_p \equiv m_p/m_{tot}$, respectively. 
The orbital frequency $\omega$ for the triangle 
satisfies 
$\omega^2 = m_{tot}/a^3$.  

\begin{figure}[t]
\includegraphics[width=10cm]{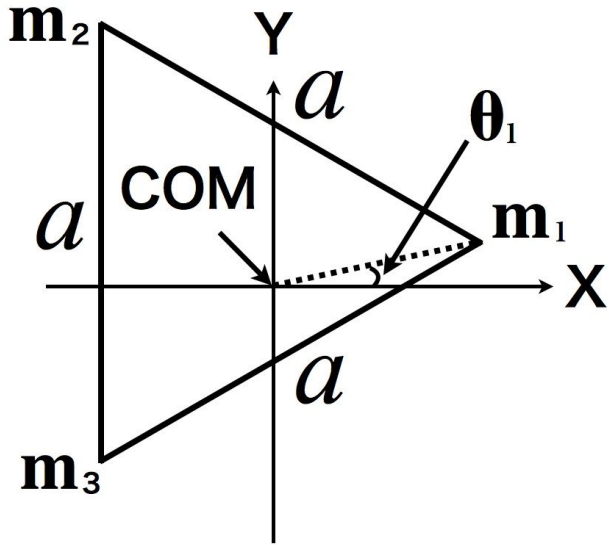}
\caption{
Definition of $\theta_p$ 
in the Lagrange's triangle solution.  
The angle $\theta_p$ is measured from $X$-axis 
to the direction of each mass at the initial time.  
}
\label{f3}
\end{figure}

Henceforth, it is convenient to employ the COM coordinates 
$(X, Y)$ that can be obtained by a translation from $(x, y)$. 
In the COM coordinates, the location of each mass at any time 
is expressed as 
$\mbox{\boldmath $X$}_p = 
a_p (\cos(\omega t+\theta_p), \sin(\omega t+\theta_p))$, 
where $a_p$ is defined as 
$a_1=\sqrt{x_{COM}^2+y_{COM}^2}$, 
$a_2=\sqrt{(3^{1/2}a/2-x_{COM})^2+(a/2-y_{COM})^2}$, 
and 
$a_3=\sqrt{x_{COM}^2+(a-y_{COM})^2}$, respectively, 
and $\theta_p$ denotes the angle between the new $X$-axis 
and the direction of each mass at $t=0$ 
(See Fig. $\ref{f3}$). 

By using the standard quadrupole formula, 
the energy loss rate for the Lagrange's orbit 
is expressed as 
\begin{eqnarray}
\frac{dE}{dt}
=
\frac{32}{5} m_{tot}^2 \omega^6 
\left[ 
\left( \sum_{p=1}^3 {\nu}_p a_p^2 \right)^2 
-4 \sum_{p < q} {\nu}_p {\nu}_q a_p^2 a_q^2 
\sin^2(\theta_p-\theta_q) 
\right] . 
\label{dEdt1}
\end{eqnarray}

The equation of motion for each body is rewritten 
in an effective one-body form as \cite{Danby} 
$d^2 \mbox{\boldmath $X$}_p/dt^2 
= -M_p \mbox{\boldmath $X$}_p / |\mbox{\boldmath $X$}_p|^3$, 
where we define the {\it effective} mass as 
\begin{equation}
M_p = m_{tot} 
\left(\sum_{q \neq p} \nu_q^2 + \sum_{q, r \neq p} \nu_q \nu_r/2 
\right)^{3/2} . 
\label{Mi}
\end{equation}
The orbital frequency is the same for each body, which provides 
an identity as $M_p/a_p^3 = \omega^2$ from 
the above effective one-body equation of motion. 
One can reexpress $a_p$ as 
$a_p = (M_p/m_{tot})^{1/3} a$ 
in terms of $M_p$ 
because $\omega^2 = m_{tot}/a^3$. 

For the triangle solution, 
we obtain the sum of the Newtonian kinetic and potential energy as 
\begin{eqnarray}
E_{tot} 
&=& 
-\frac{m_{tot}^2}{2a} 
\left[ 
\sum_{p \neq q} \nu_p \nu_q 
- \sum_p \nu_p \left(\frac{M_p}{m_{tot}}\right)^{2/3} 
\right] . 
\label{Etot}
\end{eqnarray}
By assuming adiabatic changes, 
we use the energy balance between the system energy loss 
and gravitational radiation. 
We find 
\begin{eqnarray}
\frac{1}{a}\frac{da}{dt}
&=&
-\frac{64}{5}
\frac{m_{tot}^3}{a^4}
\frac{
\left\{
\sum_p \nu_p \left(\frac{M_p}{m_{tot}}\right)^{2/3} 
\right\}^2 
-2 \sum_{p \neq q} \nu_p \nu_q 
\left(\frac{M_p}{m_{tot}}\right)^{2/3} 
\left(\frac{M_q}{m_{tot}}\right)^{2/3} 
\sin^2(\theta_p - \theta_q) 
}
{
\sum_{p \neq q} \nu_p \nu_q 
- \sum_p \nu_p \left(\frac{M_p}{m_{tot}}\right)^{2/3} 
} 
,  
\label{da}
\end{eqnarray}
which provides the shrinking rate of the triangle due to 
gravitational radiation reaction. 

Since the gravitational waves frequency $f_{GW}$ 
is twice of the orbital one, we have 
$f_{GW}^2 = m_{tot}/\pi^2 a^3$. 
Therefore, $d\ln f_{GW}/dt = -(3/2) d\ln a/dt$. 
Using this in Eq. ($\ref{da}$), 
we obtain 
\begin{eqnarray}
\frac{1}{f_{GW}}\frac{df_{GW}}{dt}
&=&
\frac{96}{5} \pi^{8/3} M_{chirp}^{5/3} f_{GW}^{8/3} ,  
\label{df}
\end{eqnarray}
where we defined a chirp mass as 
\begin{eqnarray}
M_{chirp} 
&=& 
m_{tot}  
\left[ 
\frac{
\left\{
\sum_p \nu_p \left(\frac{M_p}{m_{tot}}\right)^{2/3} 
\right\}^2 
-2 \sum_{p \neq q} \nu_p \nu_q 
\left(\frac{M_p}{m_{tot}}\right)^{2/3} 
\left(\frac{M_q}{m_{tot}}\right)^{2/3} 
\sin^2(\theta_p - \theta_q) 
}
{
\sum_{p \neq q} \nu_p \nu_q 
- \sum_p \nu_p \left(\frac{M_p}{m_{tot}}\right)^{2/3} 
}
\right]^{3/5} . 
\label{Mchirp}
\end{eqnarray} 
It is worthwhile to mention that 
the frequency sweep for the triple system 
can take the same form as that for binaries. 
One can show that Eq. ($\ref{Mchirp}$) recovers 
the binary chirp mass in the limit of $m_3 \to 0$. 

Equation ($\ref{df}$) suggests that 
we cannot distinguish two cases of the binary  
and triple systems by using only the quadrupolar parts
even if the frequency sweep is observed.

\begin{figure}[t]
\includegraphics[width=12cm]{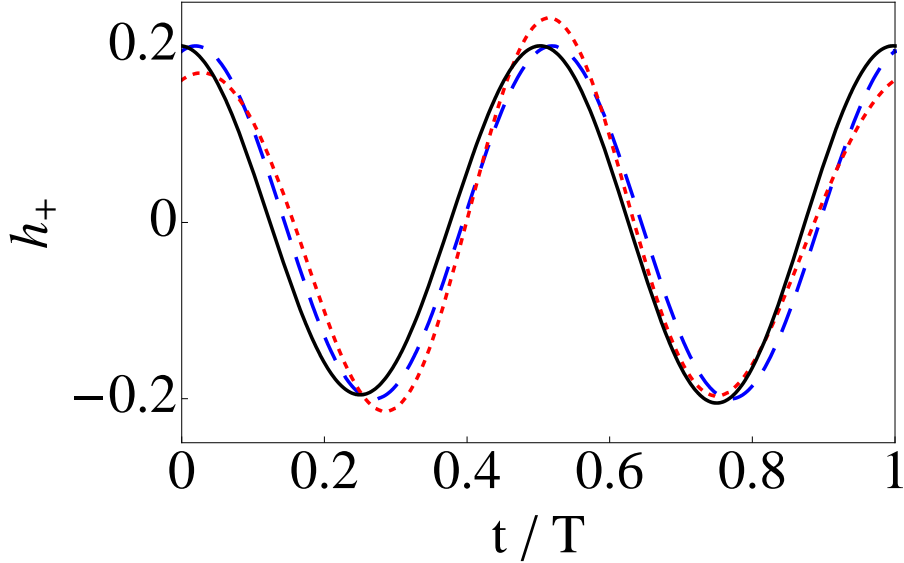}
\caption{
Gravitational waveforms in arbitrary units 
for a binary (solid black curve) 
with $m_1:m_2 = 2:3$ 
and 
a Lagrange solution (dotted red one) 
with $m_1:m_2:m_3 = 1:2:3$, 
where both the quadrupole and octupole parts are included. 
As a reference, we give the quadrupolar waveforms from 
the same sources (dashed blue). 
We assume $a=100 m_{tot}$ in order to exaggerate a correction 
by the octupole (nearly ten percents expected in this figure). 
The direction to the observer is along $x$-axis. 
One can see that the dashed blue curve will overlap with 
the solid black one after they are shifted 
by choosing the initial phase. 
This coincidence is because the octupolar waves for the binary case 
are proportional to the mass difference \cite{BS} 
and thus relatively small in this figure. 
}
\label{f4}
\end{figure}

\noindent \emph{Octupole waveforms.---}
In a wave zone, the gravitational waves denoted by $h^{TT}_{ij}$ 
can be expressed asymptotically in multipolar expansions \cite{Thorne}.  
The ratio of the octupole part to the quadrupole one 
is of the order of $v/c$, where $v$ is a typical velocity of 
the matter. 
For instance, it is about ten percents if $a=100 m_{tot}$, 
which is assumed in order to exaggerate 
the octupole correction in Fig. $\ref{f4}$. 

After straightforward calculations, one can obtain 
an expression of octupolar parts of the gravitational waves 
that are generated by the three-body system for 
the Lagrange's solution with arbitrary mass ratio. 
For instance, one of the relevant octupole moments is expressed as 
\begin{eqnarray}
I_{xxy}
=
\frac{1}{20}
\sum_{p=1}^3 m_p |\mbox{\boldmath $X$}_p|^3  \sin(\omega t +\theta_p)
-\frac{1}{4} \sum_{p=1}^3 m_p |\mbox{\boldmath $X$}_p|^3  
\cos 3(\omega t +\theta_p) . 
\label{octupole}
\end{eqnarray} 
$I_{xyy}$ can be obtained by interchanges as 
$x \leftrightarrow y$ and $\sin \leftrightarrow \cos$. 
By using such analytic expressions, one can obtain 
the octupole contributions to waveforms. 

It should be noted that no octupole radiation is emitted 
along the orbital axis for any planar motions. 
Let us take the observational direction along $x$-axis. 
Then, we have only $+$ mode without $\times$ mode. 
Figure $\ref{f4}$ shows that a difference 
between the waveforms (one by the binary 
and the other by the triplet) 
comes up at the octupole order. 
The octupole radiation amplitude by binaries is proportional 
to the mass difference \cite{BS}. 
On the other hand, the octupole radiation exists 
for triangles even if they are all equal masses. 
Cases of various mass ratios and 
observational directions are a topic of future study. 

\noindent \emph{Conclusion.--- } 
In summary, we have examined 
different numbers of self-gravitating particles 
in gravitational waves astronomy. 
In order to track the evolution of the similar waveforms from 
the two-body and three-body systems,   
we have defined a chirp mass to the three-body case. 
We have shown that the waveforms at the quadrupole level 
cannot distinguish the sources 
even with observing frequency sweep. 
Our example suggests that theoretical waveforms 
including higher multipole parts 
will be important for classification of such similar imprints. 
Higher post-Newtonian corrections both to the waveforms 
and to the motion of bodies should be incorporated. 
This is a topic of future study. 
In particular, the stability of the Lagrange orbit due to 
general relativistic effects is poorly understood. 

It is conjectured by induction from our result that classification of 
N (or less) particles producing (nearly) the same waveforms 
requires inclusions of 
the $\ell$-th multipole part with $\ell \leq N$. 
Cases of $\ell < N$ are realized for instance by 
the criss-cross and figure-eight. 
Proving (or disproving) this conjecture is left as future work. 


\section*{Acknowledgment} 
We are grateful to P. Hogan for useful comments 
on the manuscript. 
We would like to thank Y. Kojima 
for useful conversations 
at the JGRG18 workshop in Hiroshima.

\end{document}